\documentclass[aps,prl,amsmath,floats,floatfix, twocolumn,
superscriptaddress,nofootinbib,showpacs,longbibliography]{revtex4-1}
\usepackage{ifpdf}
\ifpdf
\pdfoutput=1  
\fi

\usepackage[T1]{fontenc}
\usepackage[utf8]{inputenc}
\usepackage{lmodern}

\usepackage{verbatim}
\usepackage{placeins}

\usepackage[dvipsnames, usenames]{xcolor}
\definecolor{linkcolor}{rgb}{0.0,0.3,0.5}
\usepackage[hypertexnames=false, unicode, colorlinks=true, linkcolor=linkcolor,
citecolor=linkcolor, filecolor=linkcolor,urlcolor=linkcolor,
pdfusetitle]{hyperref}

\usepackage[all]{hypcap}
\usepackage{graphicx}
\usepackage{xspace}
\usepackage{amssymb}
\usepackage[normalem]{ulem} 
\usepackage{bm} 

\usepackage{microtype}

\usepackage[english]{babel}
\usepackage{blindtext}

\DeclareMathAlphabet{\mathpzc}{OT1}{pzc}{m}{it}

\usepackage{orcidlink}

\begin{document}

\title{Eccentric binary black holes:\\ A new framework for numerical relativity waveform surrogates}

\newcommand{\AEI}{
    Max Planck Institute for Gravitational Physics
    (Albert Einstein Institute),
    Am M{\"u}hlenberg 1,
    14476 Potsdam, Germany}%
\newcommand{\BHam}{School of Physics and Astronomy and Institute for
    Gravitational Wave Astronomy, University of Birmingham, Edgbaston,
    Birmingham, B15 9TT, UK}
\newcommand{\Caltech}{Theoretical Astrophysics 350-17,
    California Institute of Technology, Pasadena, CA 91125, USA}
\newcommand{\CENTRA}{CENTRA, Departamento de F\'{\i}sica, Instituto Superior T\'ecnico, Universidade de Lisboa, Avenida Rovisco Pais 1, 1049-001 Lisboa, Portugal}
\newcommand{\Cornell}{Cornell Center for Astrophysics and Planetary Science,
    Cornell University, Ithaca, New York 14853, USA}
\newcommand{\CITA}{Canadian Institute for Theoretical
    Astrophysics, 60 St.~George Street, University of Toronto,
    Toronto, ON M5S 3H8, Canada} %
\newcommand{\JPL}{Jet Propulsion Laboratory, California Institute of Technology, Pasadena, CA
    91109, USA}
\newcommand{\KITP}{Kavli Institute for Theoretical Physics, University of California Santa Barbara, Kohn Hall, Lagoon Rd, Santa Barbara, CA 93106, USA}
\newcommand{\Manchester}{University of Manchester, Manchester, UK}
\newcommand{\MIT}{Department of Physics and MIT Kavli Institute, Cambridge, MA 02139, USA}
\newcommand{\Pullman}{Department of Physics and Astronomy, Washington State
    University, Pullman, Washington 99164, USA}
\newcommand{\UMiss}{Department of Physics and Astronomy,
    University of Mississippi, University, MS 38677, USA}
\newcommand{\Radboud}{Department of Astrophysics/IMAPP, Radboud University Nijmegen, P.O. Box
    9010, 6500 GL Nijmegen, The Netherlands}
\newcommand{\TorontoPhysics}{Department of Physics
    60 St.~George Street, University of Toronto,
    Toronto, ON M5S 3H8, Canada} %
\newcommand{\Tokyo}{Research Center for the Early Universe, University of Tokyo, Tokyo, 113-0033, Japan}%
\newcommand{\GWPAC}{Nicholas and Lee Begovich Center for Gravitational-Wave Physics and
    Astronomy, California State University Fullerton,
    Fullerton, California 92834, USA} %
\newcommand{\ChristopherNewport}{Christopher Newport University, Newport News, VA 23606, USA}
\newcommand{\UMass}{Department of Mathematics,
    Center for Scientific Computing and Data Science Research,
    University of Massachusetts, Dartmouth, MA 02747, USA
} %
\newcommand{\UTA}{Theory Group, Department of Physics, University of Texas at Austin, Austin, TX 78712, USA}
\newcommand{\Icts}{International Centre for Theoretical Sciences, Tata Institute
    of Fundamental Research, Bangalore 560089, India}
\newcommand{\Coimbra}{CFisUC, Department of Physics, University of
    Coimbra, 3004-516 Coimbra, Portugal}
\newcommand{\Balears}{Departament de Física, Universitat de les Illes Balears,
    IAC3 -- IEEC, Crta.~Valldemossa km 7.5, E-07122 Palma, Spain}
\newcommand{\KingsLondon}{Theoretical Particle Physics and Cosmology Group,
    Physics Department, King's College London, Strand, London WC2R 2LS, United
    Kingdom}
\newcommand{\VSM}{{Department of Physics, Vivekananda Satavarshiki
            Mahavidyalaya (affiliated to Vidyasagar University), Manikpara 721513, West
            Bengal, India}}
\newcommand{\Oberlin}{Department of Physics and Astronomy, Oberlin College}
\newcommand{\Wigner}{HUN-REN Wigner RCP, H-1121 Budapest, Konkoly Thege Mikl\'{o}s \'{u}t  29-33, Hungary}
\newcommand{\Illinois}{The Grainger College of Engineering, Department of Physics \& Illinois Center for Advanced Studies of the Universe, University of Illinois Urbana-Champaign, Urbana, Illinois 61801, USA}
\newcommand{\Texas}{Department of Physics and Weinberg Institute for Theoretical Physics, University of Texas at Austin, TX 78712, USA}

%
%
\address{\AEI}
\address{\UMass}
\address{\KITP}
\address{\Cornell}
\address{\GWPAC}
\address{\Illinois}
\address{\Icts}
\address{\Balears}
\address{\Texas}
\address{\CENTRA}
\address{\Caltech}
\address{\VSM}

\author{%
    Peter~James~Nee~\orcidlink{0000-0002-2362-5420}$^{*,1}$,
    Adhrit~Ravichandran~\orcidlink{0000-0002-9589-3168}$^{2}$,
    Scott~E.~Field~\orcidlink{0000-0002-6037-3277}$^{2}$,
    Tousif~Islam~\orcidlink{0000-0002-3434-0084}$^{3}$,
    Harald~P.~Pfeiffer~\orcidlink{0000-0001-9288-519X}$^{1}$,
    Vijay~Varma~\orcidlink{0000-0002-9994-1761}$^{2}$,
    Michael~Boyle~\orcidlink{0000-0002-5075-5116}$^{4}$,
    Andrea~Ceja~\orcidlink{0000-0002-1681-7299}$^{5}$,
    Noora~Ghadiri$^{5,6}$,
    Lawrence~E.~Kidder~\orcidlink{0000-0001-5392-7342}$^{4}$,
    Prayush~Kumar~\orcidlink{0000-0001-5523-4603}$^{7}$,
    Akash~Maurya~\orcidlink{0009-0006-9399-9168}$^{7}$,
    Marlo~Morales~\orcidlink{0000-0002-0593-4318}$^{5}$,
    Antoni~Ramos-Buades~\orcidlink{0000-0002-6874-7421}$^{8}$,
    Abhishek~Ravishankar~\orcidlink{0009-0006-6519-8996}$^{2}$,
    Katie~Rink~\orcidlink{0000-0002-1494-3494}$^{9}$,
    Hannes~R.~R\"uter~\orcidlink{0000-0002-3442-5360}$^{10}$,
    Mark~A.~Scheel~\orcidlink{0000-0001-6656-9134}$^{11}$,
    Md~Arif~Shaikh~\orcidlink{0000-0003-0826-6164}$^{12}$
    and
    Daniel~Tellez~\orcidlink{0009-0008-7784-2528}$^{5}$
}

\hypersetup{pdfauthor={Nee et al.}}

\date{\today}

\begin{abstract}
    Mounting evidence indicates that some of the gravitational wave signals observed by the LIGO/Virgo/KAGRA observatories might arise from eccentric compact object binaries, increasing the urgency for accurate waveform models for such systems.  While for non-eccentric binaries, surrogate models are efficient and accurate, the additional features due to eccentricity have posed a challenge.
    In this letter, we present a novel
    method for decomposing eccentric numerical relativity waveforms which makes
    them amenable to surrogate modelling techniques.
    We parameterize the inspiral in the radial phase domain, factoring
    out eccentricity-induced dephasing and thus enhancing compressibility and accuracy.  This is combined with a second surrogate for the merger-ringdown in the time-domain and a novel technique to take advantage of the approximate periodicity with radial oscillations during the inspiral.
    We
    apply this procedure to the $(2,2)$ mode for non-spinning black hole
    binaries, and demonstrate that the resulting surrogate, \texttt{NRSurE\_q4NoSpin\_22}, is able to faithfully
    reproduce the underlying numerical relativity waveforms, with maximum
    mismatches of $5\times10^{-4}$ and median mismatches of $2\times10^{-5}$.
    This technique paves the way for high-accuracy parameter estimation with
    eccentric models, a key ingredient for astrophysical inference and
    tests of general relativity.
\end{abstract}

\maketitle

\renewcommand{\thefootnote}{}
\footnotetext{$^*$ Corresponding author; \href{mailto:peter.nee@aei.mpg.de}{peter.nee@aei.mpg.de}}
\renewcommand{\thefootnote}{\arabic{footnote}}
\setcounter{footnote}{0} 

\paragraph{\textbf{Introduction:}}

The detection of gravitational waves from binary black hole coalescences has transitioned from rare, milestone events~\cite{Abbott:2016blz, Abbott:2016nmj, Abbott:2017vtc} to routine occurrences~\cite{LIGOScientific:2018mvr, LIGOScientific:2020kqk, LIGOScientific:2021usb,nitz20234,LIGOScientific:2025slb}. Continued improvements in detector sensitivity leading to next generation detectors~\cite{Barsotti:2018, LIGOScientific:2016wof, Punturo:2010zz,Maggiore:2019uih,Reitze:2019iox,LISA:2017pwj, Flaminio:2020lqk, Capote:2024rmo, Fritschel:2023} will increase the frequency of observations, and also raises the typical signal-to-noise ratio of louder signals.  This affords more opportunities to detect binaries from exotic configurations (like high mass-ratio~\cite{LIGOScientific:2020zkf} or a BH in the lower mass-gap~\cite{LIGOScientific:2024elc}), but will require ever more accurate waveform models covering an ever larger portion of parameter space to avoid systematic biases~\cite{Dhani:2024jja}.

Of particular interest is the detection of a system with non-negligible orbital eccentricity.  A binary system in isolation will radiate away eccentricity throughout its inspiral~\cite{peters1964gravitational, PhysRev.131.435}, and is expected to be a quasi-circular inspiral when it enters the frequency band of ground-based detectors. However, there are several mechanisms that could result in binaries being observed with measurable eccentricity; these include dynamical capture scenarios~\cite{Gondan:2018khr, Gondan:2020svr}, binaries resulting from triple interactions~\cite{Antonini:2017ash,Fragione:2020nib,Silsbee:2016djf,Arca-Sedda:2018qgq,Vigna-Gomez:2020fvw}, as well as binary-binary~\cite{Zevin:2018kzq, Heinze:2025usf} or binary-single~\cite{Samsing:2017rat, Samsing:2013kua, Heinze:2025usf} interactions. Thus, the observation of eccentric binaries would provide us with new insights into the formation of such systems, as well as the astrophysical environments in which they form.

The relaxation of the quasi-circular restriction for binary systems enlarges the parameter space by two dimensions, and introduces a new
characteristic frequency to the gravitational wave signal (corresponding to the
radial motion of the binary). Recently, there has been a significant effort in
the development of waveform models incorporating this additional
physics~\cite{Ramos-Buades:2023yhy, Gamboa:2024hli, Liu:2023ldr, Nagar:2024dzj,
    Gamba:2024cvy, Paul:2024ujx, Planas:2025feq, Albanesi:2025txj, Gamboa:2024imd,
    Morras:2025nbp, Henry:2023tka, Konigsdorffer:2006zt, Morras:2025nlp,
    Setyawati:2021gom, Chattaraj:2022tay, Manna:2024ycx, Islam:2024rhm,
    Islam:2024bza, Islam:2024zqo}. With these models in hand, several works have
begun to perform searches for eccentric signals in data provided by LIGO, Virgo,
and KAGRA~\cite{Wang:2021qsu,Nitz:2019spj,LIGOScientific:2019dag,LIGOScientific:2023lpe,Phukon:2024amh},
as well as re-analyzing previously detected events to search for
eccentricity~\cite{Morras:2025xfu,Gupte:2024jfe, Clarke:2022fma,
    Ramos-Buades:2023yhy, Planas:2025jny, Kacanja:2025kpr, Jan:2025fps}. However, there are caveats; the increased
computational cost of these models, as well as the additional dimensionality
introduced by eccentricity, currently prohibits the usage of these models in
routine parameter estimation pipelines~\cite{Kacanja:2025kpr,Gupte:2024jfe}. Along with this, these models do not
include complete eccentric corrections to the merger and ringdown phases, and so
their ability to capture eccentric signatures for very high-mass systems is
limited.

A natural candidate to alleviate these issues is the surrogate family of
waveform models~\cite{Field:2013cfa, Varma:2019csw, Varma:2018mmi,
    Blackman:2017pcm, Blackman:2015pia, PhysRevD.103.064022, Islam:2022laz,
    Gadre:2022sed}. Surrogates rely on a data-driven, reduced order modelling
approach, where one interpolates between a set of basis waveforms. A typical
choice of basis waveforms are those resulting from numerical relativity (NR)
simulations~\cite{Baumgarte:2010ndz,Pretorius:2005gq,Campanelli:2005dd,Baker:2005vv,Mroue:2013xna,Boyle:2019kee,Scheel:2025jct}. NR involves fully simulating Einstein's
equations with no underlying approximations~\cite{Baumgarte:2010ndz}, and so captures all of the relevant
physics of the system. These waveforms, while costly, serve as the ``ground
truth'' for evaluating the precision of approximate waveform models, and are
routinely used for calibration and improvement of said models (e.g.~\cite{Pompili:2023tna,Pompili:2024yec,Pratten:2020fqn}). Surrogate models,
when trained directly on NR waveforms, are able to faithfully reproduce the
underlying waveforms to within their intrinsic errors, providing us with a model
that is both fast and accurate.

However, building surrogate models for eccentric binary systems has proven
challenging~\cite{Islam:2021mha, Islam:2025llx}, because of two main issues: 1)
effectively covering the eccentric parameter space with numerical relativity
simulations is non-trivial, and 2) surrogate modelling techniques developed for
quasi-circular binaries show unacceptable performance when applied to eccentric
systems~\cite{Islam:2025llx, Shi:2024age}. The first issue arises from sensitivity to initial conditions, and was
alleviated in recent work~\cite{Pfeiffer:2007yz, Habib:2024soh, Knapp:2024yww, Ramos-Buades:2022lgf, Nee:2025zdy}.

\begin{figure*}
    \includegraphics[width=\linewidth,trim=5 10 5 5,clip=true]{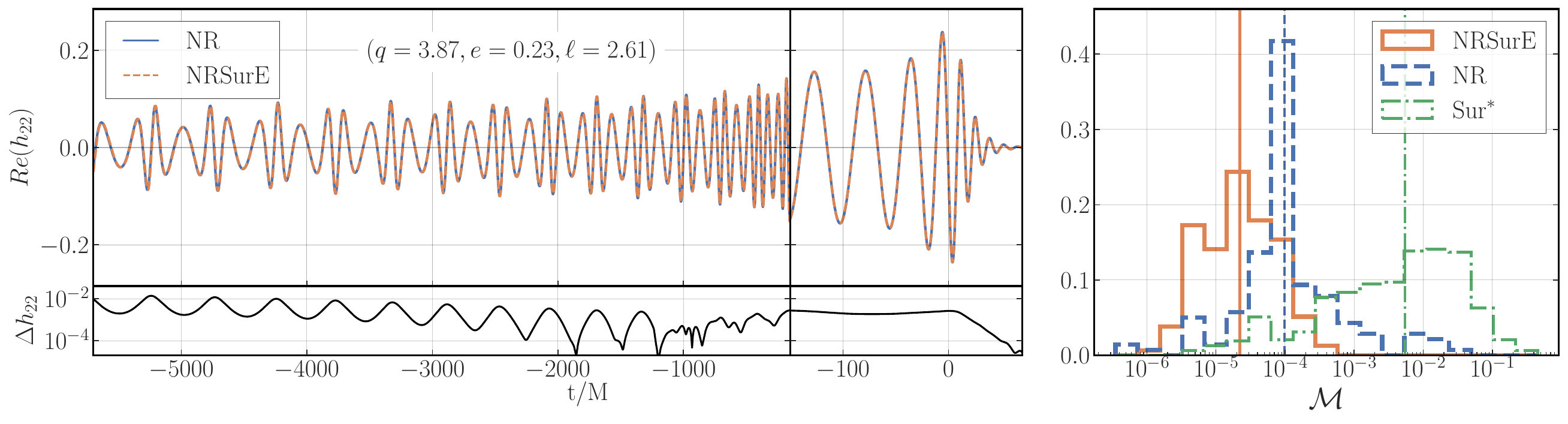}
    \caption{\textit{Demonstration of surrogate using new waveform decomposition.} \textbf{Left:} Worst validation case waveform, with mismatch of $5\times10^{-4}$. The dashed orange line is the surrogate evaluation, while the solid blue line is the corresponding NR waveform. The bottom panel shows the absolute waveform error $\Delta h_{22} = |h_{22}^{\rm Sur} - h_{22}^{\rm NR}|$. \textbf{Right:} Comparison of surrogate mismatches with NR mismatches. The solid orange lines correspond to the leave-eight-out cross validation mismatches for the surrogate, while the blue dashed curves correspond to the mismatches between the two highest resolutions of each NR waveform.  Also plotted are mismatches for a surrogate constructed using modern methods for quasi-circular systems. Vertical lines indicate medians.
    }
    \label{fig:validation}
\end{figure*}

This \textit{Letter} addresses the second issue, the ineffectualness
of quasi-circular surrogate approaches for eccentric binaries.  We present a novel strategy to decomposing eccentric numerical
relativity waveforms such that they are well-suited for surrogate modelling, and
we introduce new techniques to enforce periodicity of the radial phase. Our
approach, involving a \textit{radial-phase} re-parameterization of the system,
simplifies the data by removing the incoherence in the additional features
introduced by orbital eccentricity. We apply this procedure to the $(2,2)$ mode
for a set of $156$ non-spinning numerical relativity simulations, and show
that the resulting surrogate model faithfully reproduces the
underlying waveforms to within their intrinsic errors, as is demonstrated in
Fig.~\ref{fig:validation}.

\paragraph{\textbf{Problem outline:}}
The effectiveness of surrogate modelling relies on finding a highly amenable decomposition of the underlying waveforms. The data pieces that are chosen to be modelled should have a minimal number of features, such that they are well-described by a small number of basis functions chosen to span the waveform space efficiently. Along with this, the variation of these features across parameter space should be as simple as possible, enabling accurate regression to new parameters with only a small number of training points.
For a more complete description of the surrogate construction process, we refer the reader to~\cite{Field:2013cfa, Blackman:2017pcm}.

\begin{figure}
    \includegraphics[width=\linewidth,clip=true,trim=15 15 5 10]{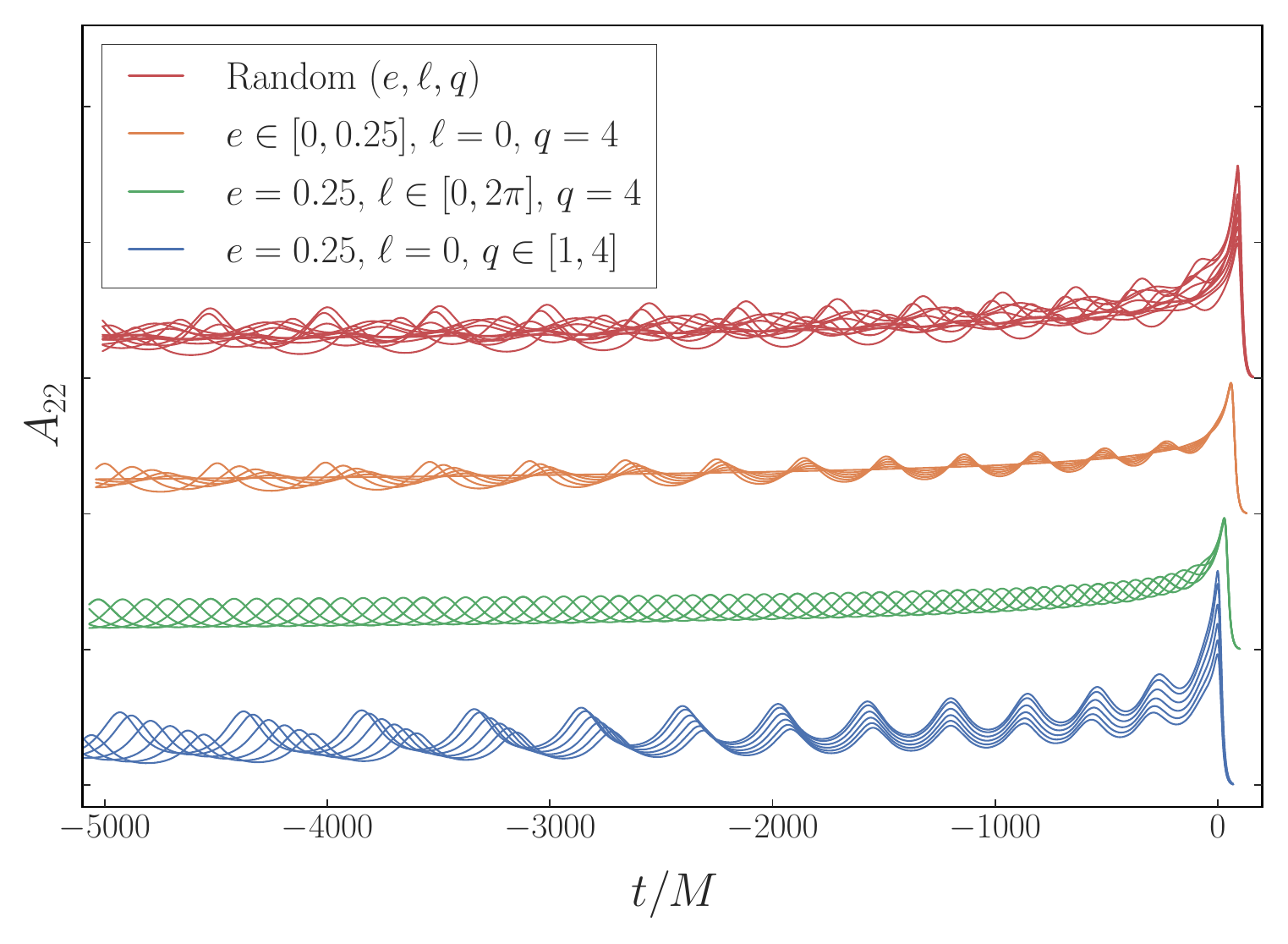}
    \caption{\textit{Incompressibility of data pieces.}  We show the amplitude of the $(2,2)$ mode $A_{22}(t)$ for sequences of waveforms generated using the surrogate model. From top to bottom, we first vary all 3 parameters (eccentricity $e$, mean anomaly $\ell$, and mass ratio $q$) together, then vary $e$, $\ell$, and $q$ individually while keeping the other parameters constant. Each sequence of waveforms is offset vertically and horizontally for clarity.
    }
    \label{fig:problem}
\end{figure}

As is standard, we first decompose the complex waveform $h=h_+-ih_\times$ into a
sum of spin-weighted spherical harmonics $h_{\ell m}$. We restrict to non-spinning
systems, and consider only the dominant $(2,2)$ mode, which we further decompose
into an amplitude and phase $h_{22}(t)=A_{22}(t)e^{-i\phi_{22}(t)}$. We choose
$t=0$ to correspond to the peak of $\Sigma_{m=-2}^2(A_{2m})^2$.

Figure~\ref{fig:problem} demonstrates the difficulties that arise when one attempts to apply standard surrogate techniques to eccentric waveforms. Historically, the construction of surrogates for quasi-circular, non-precessing systems has involved modelling $A_{22}(t)$ and $\phi_{22}(t)$ directly as functions of time~\cite{Field:2013cfa}. When considering eccentric systems, the main difficulty arises from the additional features and timescales introduced by orbital eccentricity; namely the oscillations on the orbital time scale in the amplitude and phase. As can be seen in Fig.~\ref{fig:problem}, these oscillations move in and out of phase across parameter space, rendering the data pieces poorly suited to efficient modeling. In fact, one of the intrinsic parameters of the system (the mean anomaly $\ell$) determines the relative phasing of these oscillations, while the remaining parameters (for this study, the mass-ratio $q$ and eccentricity $e$) determine the radial frequency of the system, meaning that even systems that are initially aligned can de-phase. Similar features are present in $\phi_{22}(t)$.

This situation is reminiscent of issues encountered for precessing-spin systems; if one were to naively attempt to model the gravitational waveform in the inertial frame (i.e., directly model $A_{22}(t)$ and $\phi_{22}(t)$), oscillations and mode-mixing introduced by the precession of the orbital plane would again render the data pieces difficult to model. As such, precessing surrogates are constructed in a frame that follows the precession of the orbital plane, such that precession-induced oscillations are minimized.

While this co-precessing-frame method has been highly effective for modelling precession-induced effects, no analogous frame transformation exists for simplifying the structure of eccentricity induced oscillations. As a result, we must pursue a different approach for such systems. The primary issue
highlighted by Fig.~\ref{fig:problem} is the temporal incoherence of the
features across the entire parameter space. To remove this incoherence, we will
extract the radial phasing throughout the inspiral, and use this to construct a
surrogate of the waveform as a function of the \textit{radial phase}, as opposed
to time. Near merger, we can no longer reliably define a radial phase, and so a
different approach is required. To this end, we construct a
separate, short surrogate model for only the final $\sim2$ orbits, as well as
merger and ringdown, using standard time-domain surrogate
techniques~\cite{Varma:2019csw}. Finally, we will align and stitch these
surrogates together, ultimately producing full inspiral-merger-ringdown (IMR)
waveforms for non-spinning systems.


\paragraph{\textbf{Inspiral:}}

\begin{figure}
    \includegraphics[width=\linewidth,trim=5 8 0 5,clip=true]{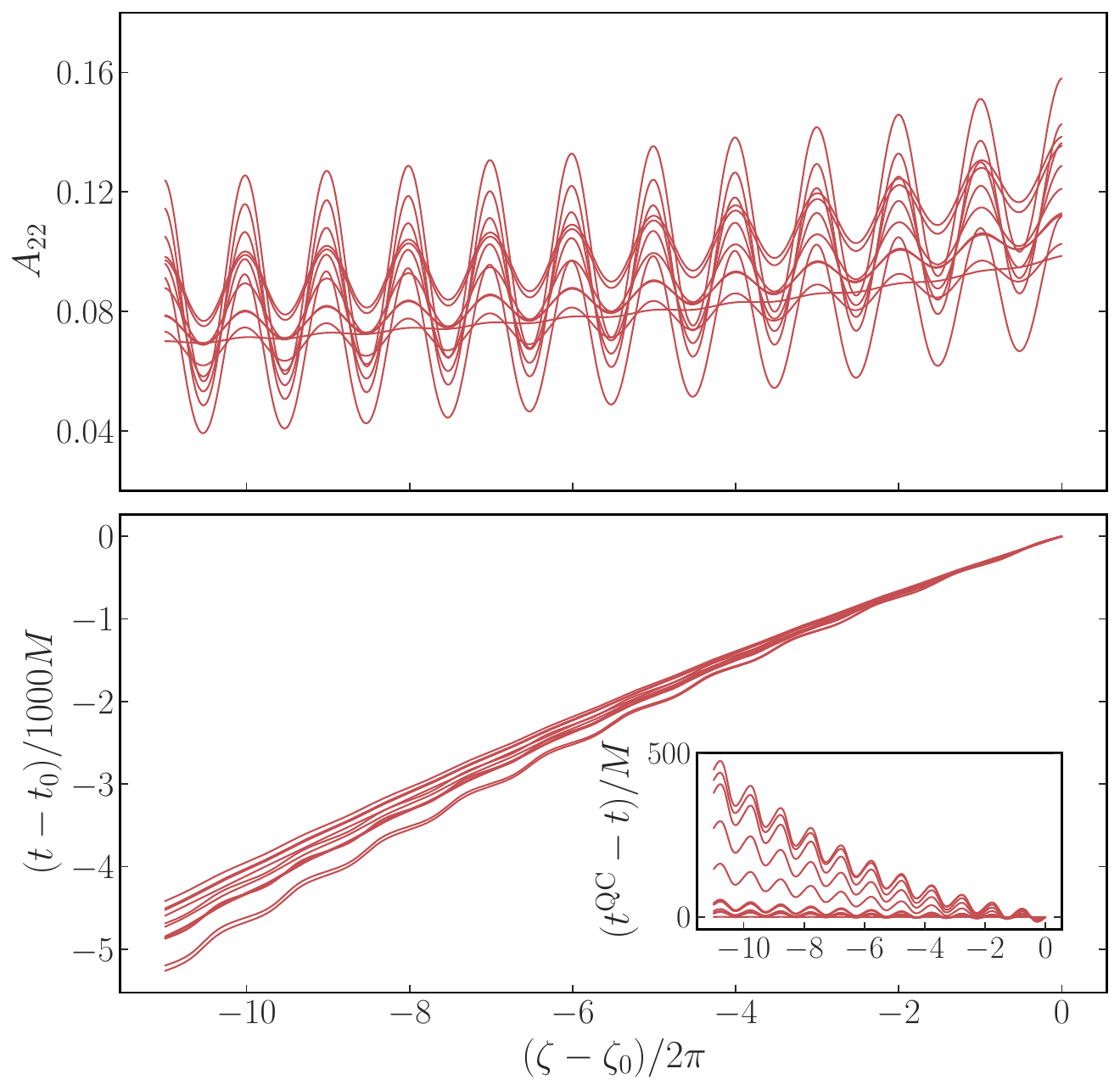}
    \caption{\textit{Example inspiral waveform decomposition.} Plotted are the
    $A_{22}(\zeta)$ and $t(\zeta)$ data pieces used for construction of the
    inspiral surrogate, for the same parameters as the top
    family in Fig.~\ref{fig:problem}. Varying radial frequencies are captured by
    $t(\zeta)$, while relative phasing is handled by how the inspiral is
    stitched to merger. Here $t_0$ ($\zeta_0$) corresponds to the time (relativistic anomaly) at the final periastron passages considered for the inspiral surrogate.  The inset of the bottom panel shows the
    difference $t(\zeta)-t^{\rm QC}(\zeta)$, where the time $t^{\rm QC}(\zeta)$  for a quasi-circular inspiral is obtained by evaluating the surrogate for the same $(q,\ell_{-1200M})$, but with $e_{-3000M}$ set to zero.
    }
    \label{fig:data}
\end{figure}

In order to re-parameterize the system as a function of radial phase, we need an approach to extract a radial phase from numerical relativity simulations. Several definitions of radial phase could be chosen: for this work, we employ the relativistic anomaly~\cite{Gamboa:2024imd}, which we obtain by solving the following quasi-Keplerian equations:
\begin{subequations}
    \label{eq:PNQK}
    \begin{align}
        \dot{e}     & = -\frac{\nu e x^4}{M} \left[\frac{\left(121 e^2+304\right)}{15 \left(1-e^2\right)^{5/2}} + \text{3PN terms}\right], \label{eq:e} \\
        \dot{\zeta} & = \frac{x^{3/2}}{M}\left[\frac{ (1+e \cos\zeta)^2}{\left(1-e^2\right)^{3/2}} + \text{3PN terms}\right], \label{eq:zeta}           \\
        x           & = (M \Omega)^{2/3} \left[\frac{\left(1-e^2\right) }{(1+e \cos\zeta)^{4/3}} + \text{3PN terms}\right], \label{eq:xOmega}
    \end{align}
\end{subequations}
where $e$ corresponds to the quasi-Keplerian eccentricity, $\zeta$ the
relativistic anomaly, $\nu=m_1m_2/M^2$ the symmetric mass-ratio, $M$ the total
mass of the binary, $\Omega$ is the instantaneous orbital frequency of the
binary, and $x$ is the dimensionless orbit-averaged orbital frequency. The full
$3$PN equations can be found in~\cite{Gamboa:2024hli}. This radial phase is
desirable, as (unlike the mean anomaly used in previous work~\cite{Islam:2021mha})
it is both a smooth function of time, and has a well-defined limit
for $e\rightarrow 0$. These equations are currently derived up to 3PN for aligned
spin systems, and up to 2PN for generic systems.

To solve these equations, one must supply $(q, \chi_i, \Omega(t), e_0,
    \zeta_0)$. Since we are restricted to non-spinning systems, we set $\chi_i=0$,
and compute the mass-ratio $q$ using the Christodolou mass computed from the apparent horizons~\cite{Scheel:2025jct}. For $\Omega(t)$, we use
the orbital frequency computed using the coordinate center's of the individual
black-holes.

For $(e_0,\zeta_0)$, we employ the fact that the mean anomaly $\ell_{\rm gw}$ (as defined in~\cite{Shaikh:2023ypz}) and relativistic anomaly should agree at periastron passages $\ell_{\rm gw}=\zeta=0$, and so extract the eccentricity $e_{\rm gw}$ (as defined in~\cite{Ramos-Buades:2023ehm}) at the first available periastron passage using the python package \texttt{gw\_eccentricity}~\cite{Shaikh:2023ypz}. While $e_{\rm gw}\neq e$, we find that it is sufficiently close for our purposes. A future improvement could involve an optimization procedure to determine $e_0$, using $e_{\rm gw}$ as an initial guess. We then integrate Eqs.~\eqref{eq:PNQK} throughout the inspiral to obtain $\zeta(t)$.

While our use of the trajectories introduces gauge dependent quantities, we stress that this is just an internal reparameterization to aid the surrogate modelling, similar to the use of local values of the spins (which are evolved through merger via PN spin evolution equations~\cite{Gerosa:2016sys}) for constructing fits for precessing systems. Any gauge dependence is ultimately removed when we construct the time domain waveform. Another point is that we can only extract the radial phase for as long as the individual
black hole positions are tracked, and even then the
accuracy of the PN expansion used in Eqs.~\eqref{eq:PNQK} will break down if
pushed too far into the strong field regime. Therefore, we will use this
approach to construct a surrogate model only for the inspiral portion of the
waveform.

Our goal is to generate an ``inspiral'' surrogate that is applicable for times
up to $t=-1200M$. To this end, we identify the first periastron passage that
occurs after $t=-1200M$ for each waveform in the dataset, and restrict the $\zeta$-parameterisation up to this
point. Due to length limitations of the underlying NR waveforms, we consider the $11$ radial periods immediately before this for
modelling. In addition to modelling $A_{22}(\zeta)$ and $\phi_{22}(\zeta)$, we also model $t(\zeta)$, so that we can transform the surrogate evaluation back to the time domain. Finally, for only the inspiral surrogate, we perform a rotation and time-shift to each waveform,
such that $t=0$ and $\phi_{22}=0$ at the final periastron passage, to further
simplify the data.

Figure~\ref{fig:data} presents two of the exact data pieces to be modelled. Plotted are the same $12$ simulations presented in the top sequence of Fig.~\ref{fig:problem}. As a function of $\zeta$, all of the eccentricity induced oscillations in $A_{22}(\zeta)$ and $t(\zeta)$ are are mapped to the same period, and are in phase. The varying orbital period is captured by the monotonic trend in $t$, while the relative phasing will be handled by how we stitch the inspiral surrogate to the merger. The final data piece, $\phi_{22}(\zeta)$, shows similar features to $t(\zeta)$, and so we omit it for brevity. Ultimately, data decomposed as in Fig.~\ref{fig:data} is much more amenable to surrogate modelling than that in Fig.~\ref{fig:problem}.

The final choice before building the inspiral surrogate lies in the
precise definition of parameters assigned to each waveform, such that parametric fits can be performed across the parameter space. For the eccentric parameters, we choose the mean-anomaly and eccentricity as extracted from the gravitational waveform~\cite{Shaikh:2023ypz}. We define the mean-anomaly $\ell_{-1200M}$ at time $t=-1200M$ before merger, as this time facilitates the attachment of the merger surrogate. Because the eccentricity itself is difficult to extract
sufficiently accurately close to merger, we utilize
the eccentricity at $t=-3000M$ (named $e_{-3000M}$).\footnote{Recent improvements in
    \texttt{gw\_eccentricity}~\cite{Shaikh:2025tae} or alternative waveform-based eccentricity definitions~\cite{Boschini:2024scu, Islam:2025oiv} may alleviate the need for this.} We parameterize the third dimension of the surrogate --mass-ratio $q$-- by $\log_{10}(q)$ as opposed to
$q$ directly~\cite{Varma:2018aht}.

In principle, we could now construct the inspiral surrogate.  However initial validation tests of the model indicate larger errors for $\ell_{-1200M}$ near $0$ or near $2\pi$, which arise because the surrogate process does not take into account the periodicity of the underlying waveforms: systems with mean anomaly $\ell_{-1200M}$ differing by $2\pi$ correspond to the same physical system. A similar issue would arise if, for precessing systems, one parameterized the surrogate using the angles of the individual black-hole spins.

\begin{figure}
    \includegraphics[width=\linewidth,trim=5 8 5 8,clip=true]{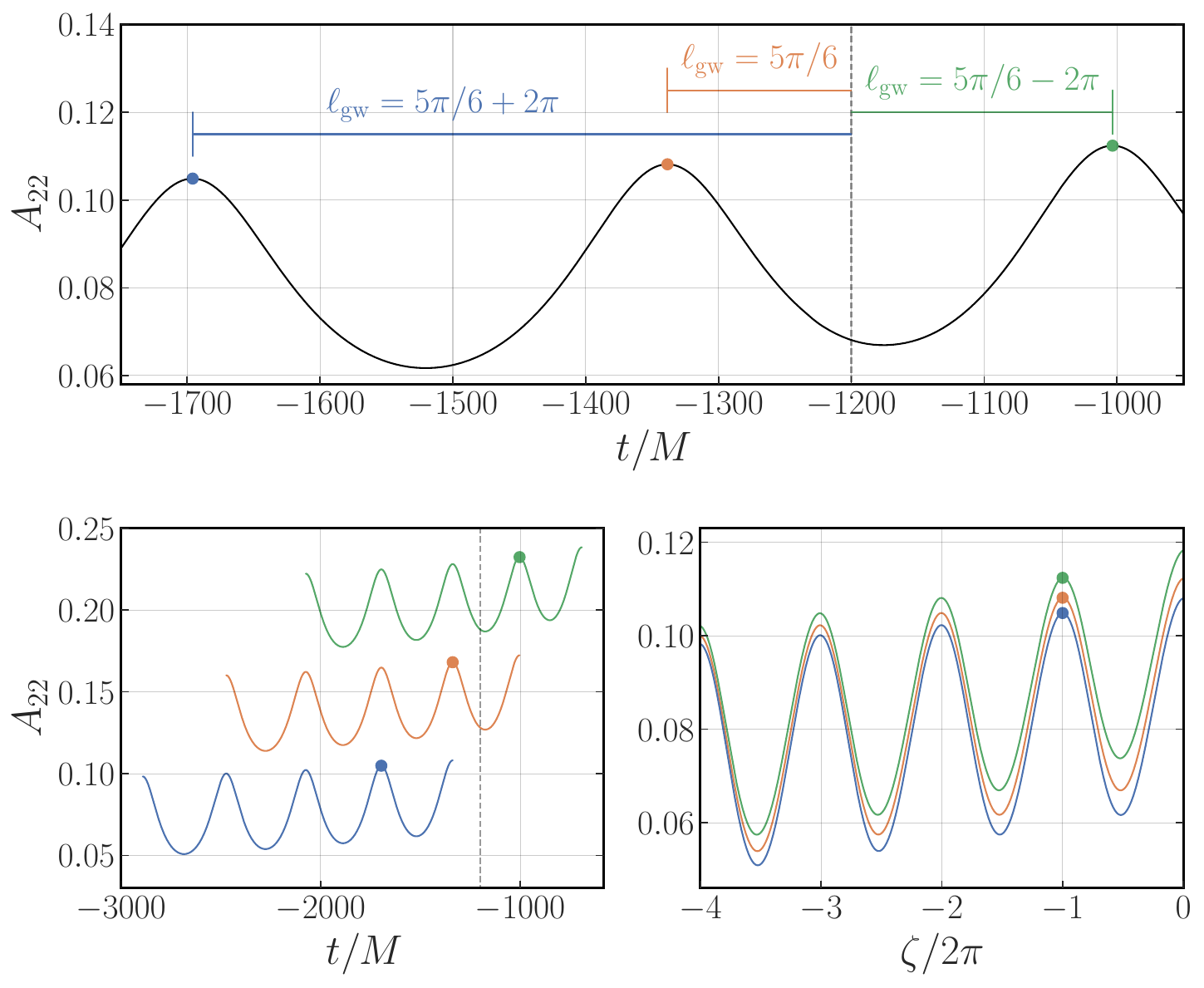}
    \caption{\textit{Demonstration of mean-anomaly periodicity in the inspiral
            surrogate.} \textbf{Top: } Illustration of how one could
        assign $\ell_{\rm gw}$ at $t=-1200M$ for three different
        reference periastron locations using one waveform. \textbf{Bottom Left: } The corresponding portion
        of waveform that would be used for each case, in time domain. For
        visibility, we vertically offset the curves from each other and
        plot only the final 4 radial periods that would be used. \textbf{Bottom Right: } The
        corresponding data pieces that are ultimately added to the surrogate
        training set, ensuring periodicity is captured for $\ell_{\rm
                gw}\in[0,2\pi]$.}
    \label{fig:duplication_inspiral}
\end{figure}

For the inspiral surrogate, building in this periodicity is somewhat subtle: the data pieces
themselves are not actually periodic in $\ell_{-1200M}$.
Instead, systems differing by
$\Delta\ell_{-1200M} = 2\pi$ correspond to
the same physical configuration shifted by one radial period, as is illustrated in Fig.~\ref{fig:duplication_inspiral}. We can exploit this pseudo-periodicity to enlarge the training set and mitigate any boundary effects, without requiring additional simulations. Specifically, we add extra training points
offset one radial period forward and backward such that the training set
contains data for $\ell_{-1200M}\in[-2\pi/3,8\pi/3]$. We emphasize that this
extension of the domain is only for improving the surrogate's parametric fits near $\ell_{-1200M}
    = 0$ and $2\pi$, and that for production waveform generation we only evaluate
$\ell_{-1200M}\in[0,2\pi]$.

\paragraph{\textbf{Full IMR Model:}}
In order to produce complete waveforms describing the coalescence of two black holes, we must combine the inspiral surrogate with a separate ``merger'' surrogate for the final stage of the binary. Here, we choose to employ a more traditional data decomposition method, modelling $A_{22}(t)$ and $\phi_{22}(t)$ directly as a functions of time in the interval $t\in[-1300M, 100M]$. While the data may not be particularly compressible in this form, the short length of the surrogate means that a reduced basis can still be produced. As well as this, since there are only $\sim2$ orbits from $t=-1300M$ to merger, very little de-phasing is able to occur, and so fits across parameter space are sufficiently simple. Finally, for defining parameters assigned to each waveform, we choose to use identical parameters ($\log(q)$, $e_{-3000M}$, and $\ell_{-1200M}$) as was done for the inspiral surrogate, such that both models can be evaluated with one set of input parameters.

Again, during validation one observes that the merger surrogate faces the same issues at $\ell_{-1200M}=0$ and $2\pi$ as the inspiral surrogate. Thus, we again extend the $\ell_{-1200M}$ domain of the surrogate. Here, this is simpler, as $A_{22}(t)$ and $\phi_{22}(t)$ will themselves be periodic in $\ell_{-1200M}$. As such, we add copies of the data to the training set where we add $\pm2\pi$ to $\ell_{-1200M}$ to extend the domain to the same as the inspiral surrogate.

At this point, we can evaluate two surrogates to get the inspiral and merger portions of a waveform. Furthermore, the choices made previously ensure that there is an overlap in the time domain between the two evaluations. Constructing a full IMR waveform then involves aligning the two evaluations in both time and phase, and smoothly transitioning between the two. In order to do this efficiently, we use the fact that the radial phasing at a particular time ($\ell_{-1200M}$) is known, and so impose that this is satisfied by both component surrogate evaluations. Further details can be found in the End Matter.

\paragraph{\textbf{Results:}}

In order to demonstrate the efficacy of this decomposition, we construct a surrogate for the $(2,2)$ mode for non-spinning eccentric systems. We use a training set of $156$ waveforms from the SXS catalog~\cite{Scheel:2025jct, Nee:2025zdy}, with identifiers SXS:BBH:[3731-3747, 3749-3776, 3778-3820, 4293-4304, 4321-4332, 4357-4381, 4442-4455]. The training set consists of mass-ratios $q\in[1,4]$ and eccentricities $e_{-3000M}\in[10^{-3}, 0.25]$. For fits across parameter space, we employ Gaussian process regression as in~\cite{Varma:2018aht, Taylor:2020bmj}. The model, which we refer to as \texttt{NRSurE\_q4NoSpin\_22}, will be made publicly available as part of the \texttt{GWSurrogate} package~\cite{Field:2025isp} in the near future.

As the inspiral portion of the surrogate is now constructed in the radial-phase domain, the waveform length (in terms of time) is no longer constant, but is instead a function of the input parameters. For this work, we obtain a waveform length of approximately $6000M$ (ranging from $5000M$ to $7000M$), corresponding to $\sim42$ gravitational wave cycles of the $(2,2)$ mode.

To validate the model, we perform a leave-eight-out cross validation study, where we divide the training set randomly into groups of 8 cases. We build a series of surrogates, each excluding one group of waveforms from their training set, which is then validated against the excluded waveforms. The effectiveness of this modelling approach is demonstrated in the right panel of Fig.~\ref{fig:validation}. Plotted are the mismatches from the leave-eight-out study, along with the mismatches obtained comparing the two highest resolutions for each system, which serves as a conservative estimate for the accuracy of the training waveforms. We see that the surrogate is able to faithfully reproduce the underlying numerical relativity waveforms, reducing the largest mismatches by several orders of magnitude, and the median mismatch by a factor of $\sim4$. This indicates that the limiting factor in the accuracy of the surrogate is likely the accuracy of the underlying numerical relativity waveforms, and not the surrogate modelling techniques (similar to what was seen in~\cite{Varma:2019csw}). As an extra illustration of the accuracy of the model, the left panel of Fig.~\ref{fig:validation} shows the worst validation case waveform, corresponding to one of the largest mass-ratios and eccentricities in the training set (i.e.\ a boundary case). The inclusion of systems with larger eccentricities and mass-ratios would likely improve the accuracy of the model for this case.

To further emphasize the importance of this framework, Fig.~\ref{fig:validation} also presents the mismatches obtained from a surrogate constructed using modern techniques for quasi-circular systems~\cite{Varma:2019csw} (indicated as Sur$^{*}$). This results in mismatches several orders of magnitude worse than the eccentric surrogate presented here. Compared to the previous eccentric NR surrogate~\cite{Islam:2021mha}, our approach is both more accurate and more robust across the parameter space, while also being readily applicable to training sets including additional intrinsic parameters.

\paragraph{\textbf{Conclusion:}}
In this \textit{Letter}, we have presented a novel approach for decomposing eccentric waveforms, enabling accurate and efficient construction of full IMR surrogate models for eccentric numerical relativity waveforms. By separating the waveform into two over-lapping regions, we are able to extract the radial phase of each simulation throughout the inspiral, and can construct a surrogate for the waveform, as well as the time-coordinate, as a function of radial phase. This removes the de-phasing of eccentricity induced oscillations, and enables reconstruction of NR waveforms to within their intrinsic errors.

While the $(2,2)$ mode predicted by the model is accurate to within the errors of the underlying NR waveforms, it is still restricted in its length. This can be alleviated by hybridization of the NR waveforms with analytical approximants~\cite{Sun:2024kmv, Varma:2018mmi}. In particular, waveform models that already solve the quasi-Keplerian equations (such as \texttt{SEOBNRv5EHM}~\cite{Gamboa:2024hli}) would be a natural choice. Regardless, by ensuring that future simulation campaigns focus on length in terms of \textit{radial periods} as opposed to frequency/time, we can ensure that the surrogate is sufficiently long for most applications.

Another application of this work is the speeding up of other waveform models for parameter estimation. Currently, most eccentric waveform models are prohibitively slow to evaluate for low mass systems, making their use in standard parameter estimation routines challenging~\cite{Kacanja:2025kpr,Gupte:2024jfe}. None of the procedure presented relies on the underlying waveforms being produced by numerical relativity simulations (and in fact, extraction of the radial phase is often easier for semi-analytic models than in numerical relativity). This has recently been demonstrated in~\cite{Maurya:2025shc}. As the exact evaluation time of the surrogate will depend on the handling of higher order modes, we refer the reader to Ref.~\cite{Adhrit:InPrep} for further details.

Finally, this work has been restricted to the non-spinning binary black hole subspace. While extension to the case of generic \textit{non-precessing} systems should be straight-forward, extension to completely generic systems remains an open problem. Several works have shown that (for quasi-circular systems), in the co-precessing frame, standard surrogate modelling techniques can be used to construct waveforms that are then ``twisted up'' to obtain the inertial frame waveform~\cite{Varma:2019csw}. This suggests that a similar procedure may be applicable for eccentric systems, and will be explored in future work.


\paragraph{\textbf{Acknowledgments:}}
The authors would like to thank Arnab Dhani, Aldo Gamboa, Marcus Haberland, Benjamin Leather, Oliver Long, Philip Lynch, and Lorenzo Pompili for useful discussions, as well as Guillermo Lara, Keefe Mitman, and Lucy Thomas for feedback on a draft of this work. P.J.N.\ would like to thank Aurora Abbondanza for one-way conversations regarding this work.
T.I. is supported in part by the National Science Foundation under Grant No.~NSF PHY-2309135 and the Simons Foundation (216179, LB).
A.~R.-B.\ is supported by the Veni research programme which is (partly) financed by the Dutch Research Council (NWO) under the grant VI.Veni.222.396; acknowledges support from the Spanish Agencia Estatal de Investigación grant PID2022-138626NB-I00 funded by MICIU/AEI/10.13039/501100011033 and the ERDF/EU; is supported by the Spanish Ministerio de Ciencia, Innovación y Universidades (Beatriz Galindo, BG23/00056) and co-financed by UIB. This material is based upon work supported by the National Science Foundation under Grants No.~PHY-2309211, No.~PHY-2309231, and No.~OAC-2209656 at Caltech; by No.~PHY-2407742, No.~PHY-2207342, and No.~OAC-2209655 at Cornell; and by No.~PHY-2208014 and AST-2219109 at Cal State Fullerton. This work was supported by the Sherman Fairchild Foundation at Caltech and Cornell, and by the Dan Black Family Trust, and Nicholas and Lee Begovich at Cal State Fullerton.
S.F.\ acknowledges support from NSF Grants No. AST-2407454 and PHY-2110496.
H.\,R.\,R.\@ acknowledges financial support provided
under the European Union's H2020 ERC Advanced Grant ``Black holes:
gravitational engines of discovery'' grant agreement no.~Gravitas–101052587.
Views and opinions expressed are however those of
the authors only and do not necessarily reflect those of the European
Union or the European Research Council.  Neither the European Union
nor the granting authority can be held responsible for them. V.V.~acknowledges support from NSF Grant No.~PHY-2309301.
This work was partly supported by UMass Dartmouth's Marine and Undersea Technology (MUST) research program funded by the Office of Naval Research (ONR) under grant No.\ N00014-23-1-2141.
The computations
presented here were conducted in the Resnick High Performance Computing Center,
a facility supported by Resnick Sustainability Institute at the California
Institute of Technology. This work used Expanse at San Diego Supercomputer
Center~\cite{Expanse}, Stampede 2 at Texas Advanced Computing Center, through
allocation PHY990002 from the Advanced Cyberinfrastructure Coordination
Ecosystem: Services \& Support (ACCESS) program, which is supported by
U.S. National Science Foundation grants \#2138259, \#2138286, \#2138307,
\#2137603, and \#2138296~\cite{NsfAccess}. The authors acknowledge the Texas
Advanced Computing Center (TACC) at The University of Texas at Austin for
providing computational resources that have contributed to the research results
reported within this paper. URL:~\url{http://www.tacc.utexas.edu}. This work
used the Extreme Science and Engineering Discovery Environment (XSEDE), which is
supported by National Science Foundation grant number ACI-1548562. Specifically,
it used the Bridges-2 system, which is supported by NSF award number
ACI-1928147, at the Pittsburgh Supercomputing Center (PSC).
Computations were performed on the Urania cluster at the
Max Planck Computing and Data Facility.
\bibliography{literature}

\FloatBarrier
\subsection*{End Matter}

\paragraph{Stitching surrogate evaluations together: } First, by construction, we know that $t=-1200M$ is within the domain of both surrogates: for the merger model, we extend as far back as $t=-1300M$, while for the inspiral model the final epoch considered is the closest periastron passage \textit{after} $t=-1200M$. This means that there will always be at least $100M$ ($-1300M<t<-1200M$) of overlap between the surrogates. However, stitching the surrogate evaluations together requires a mapping between the time co-ordinates of both surrogate evaluations (as we set $t=0$ to be the final modelled epoch in the inspiral surrogate). For clarity, we will denote the time outputted by the inspiral surrogate as $t'$. This time-shift will be a function of the input parameters of the surrogate, and so is not known apriori. However, we can utilize the fact that both surrogates are parameterized by the mean anomaly at $t=-1200M$. While $\zeta$ will not agree with $\ell_{\rm gw}$ throughout an orbit, it will agree at periastron passages, where $\ell_{\rm gw}=\zeta=2n\pi$ (modulo small de-phasing obtained from integrating Eqs.~\eqref{eq:PNQK}). We also know that, by construction, $t=-1200M$ must fall in the final period modelled by the inspiral surrogate (i.e.\ $\zeta\in[-2\pi,0]$). Thus, we can extract $t'(-2\pi)$ and $t'(0)$, and use this to determine the $t'$ that corresponds to $t=-1200M$ (labelled $t'_{-1200M}$) as
\begin{gather}
    t'_{-1200M}= \frac{\ell_{-1200M}}{2\pi}\left(t'(0)-t'(-2\pi)\right) + t'(-2\pi).
    \label{eq:time_alignment}
\end{gather}

\begin{figure}
    \includegraphics[width=\linewidth]{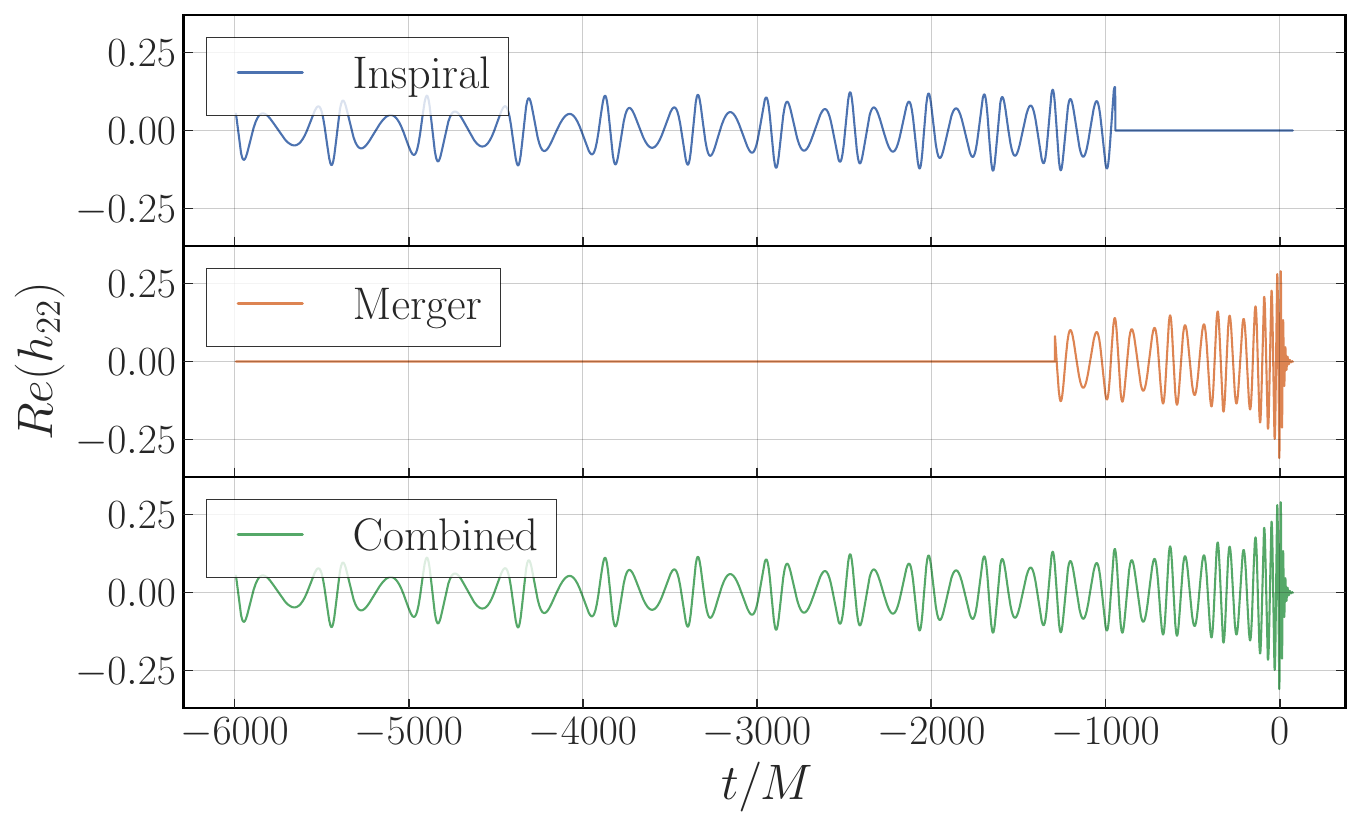}
    \caption{\textit{Example evaluations of the inspiral, merger and full IMR surrogate.} \textbf{Top:} The real part of $h_{22}$ after evaluating the surrogate, performing the time shift and rotating the waveform. \textbf{Middle:} The same for the merger surrogate evaluation. \textbf{Bottom:} The full IMR waveform after stitching both component surrogates together.}
    \label{fig:stitching}
\end{figure}

We now align the time grids by performing the shift
\begin{gather}
    t'\to t'-t'_{-1200M}-1200M.
    \label{eq:time_shift}
\end{gather}
Additionally, we rotate both waveforms such that $\phi_{22}=0$ at $t=-1200M$ to align the phases. Figure~\ref{fig:stitching} presents an example of the two surrogate evaluations after performing the time shift and rotation. The top panel shows the inspiral surrogate evaluation, while the middle panel shows the merger surrogate evaluation.

Now that the two surrogate evaluations are time and phase aligned, all that remains is to stitch them together, so that we smoothly transition from inspiral to merger. We apply Planck windowing functions centered at $t=-1230M$ with a width of $10M$ to each waveform evaluation, and sum them together. An example resulting waveform is presented in the bottom panel of Fig.~\ref{fig:stitching}.

\end{document}